\documentclass[sigconf,nonacm]{acmart}
\AtBeginDocument{%
  }

\usepackage{amsmath}
\usepackage{algorithmic}
\usepackage{graphicx}
\usepackage{textcomp}
\usepackage{enumitem}
\usepackage{xcolor}
\usepackage{booktabs}
\usepackage{multirow}
\usepackage{tikz}
\usetikzlibrary{arrows.meta, positioning, shapes.geometric}

\usepackage{colortbl}
\definecolor{g1}{RGB}{198,239,206}  
\definecolor{g2}{RGB}{217,245,223}  
\definecolor{g3}{RGB}{235,250,238}  
\definecolor{r1}{RGB}{255,199,206}  
\definecolor{r2}{RGB}{255,219,224}  
\definecolor{r3}{RGB}{255,237,240}  

\usepackage{tcolorbox}
\tcbuselibrary{breakable, skins}

\newcommand{\pipelineblock}[3]{%
  \begin{tcolorbox}[
    colback=#2!8, colframe=#2!70,
    fonttitle=\bfseries\small,
    title=#1,
    left=4pt, right=4pt, top=2pt, bottom=2pt,
    boxsep=2pt
  ]
  \ttfamily\footnotesize #3
  \end{tcolorbox}
}

\usepackage{listings}

\usepackage{tabularx}
\newcolumntype{Y}{>{\raggedright\arraybackslash}X}

\lstdefinestyle{verilog}{
    language=Verilog,
    basicstyle=\ttfamily\small,
    keywordstyle=\color{blue},
    commentstyle=\color{gray},
    stringstyle=\color{purple},
    breaklines=true,     
    frame=single,        
    captionpos=b,        
    keepspaces=true
}

\renewcommand\footnotetextcopyrightpermission[1]{}

\begin{document}

\title{VHDLSuite: Unified Pipeline for LLM VHDL Generation with Data Synthesis and Evaluation
}


\author{\large Yijun Shen$^{1}$, Minghao Shao$^{2,3}$, Yichen Zhao$^{1}$, Zhuoyan Yu$^{1}$, Boyuan Chen$^{2,3}$, Yik-Cheung Tam$^{1}$, Muhammad Shafique$^{3}$}
\affiliation{\large%
  \institution{%
    $^{1}$Center for Data Science, NYU Shanghai, China \quad
    $^{2}$NYU Tandon School of Engineering, USA \quad
    $^{3}$NYU Abu Dhabi, UAE}
  \city{}\country{}
}
\email{{ys6126, shao.minghao, yz10469, zy3077, bc3194, wilson.tam, muhammad.shafique}@nyu.edu}

\begin{abstract}
Large Language Models (LLM) have shown impressive capabilities in Register Transfer Level (RTL) code generation, particularly for Verilog. However, evaluating their performance with other Hardware Description Languages (HDL), especially VHDL, remains limited although its distinct language characteristics, such as stricter semantic rules, introduce evaluation considerations that differ from Verilog. This lack of coverage restricts fully understanding of how well current models generalize across hardware design languages with differing structures and semantics. To address this gap, we introduce VHDLSuite, a benchmark-centered infrastructure for scalable VHDL generation evaluation, integrating automated benchmark synthesis, executable validation, and multi-model diagnostic analysis. First, we propose a data pipeline that automatically converts Verilog designs and their accompanying testbenches into executable VHDL benchmark instances, followed by VUnit/GHDL-based validation to ensure each released task is compilable, runnable, and consistently checkable in the VHDL environment. Second, we introduce VHDLBench, a benchmark with over 200 VHDL problems with complete and validated testbenches across a wide range of complexity levels. Third, we extensively evaluate cutting-edge LLMs and uncover key challenges specific on LLM-aided VHDL generation. Our findings provide important insights and support future work in multi-language hardware design automation.\footnote{Our data pipeline, benchmark, and evaluation framework will be open-sourced.}
\end{abstract}

\maketitle

\section{Introduction}


Large Language Models (LLMs) have become a transformative force in hardware design automation, advancing the field across generation, optimization, and security~\cite{thakur2023autochip, wang2025netdetox, wang2025salad}. Recent advances in Register Transfer Level (RTL) code generation rely on specialized datasets and hardware-focused benchmarks~\cite{liu2024openllm}. Both general-purpose and domain-specific LLMs fine-tuned on RTL corpora now produce synthesizable hardware descriptions with promising accuracy~\cite{thakur2024verigen, liu2024rtlcoder, wei2025vericoder, deng2025scalertl}, and agentic design flows further improve output quality through iterative synthesis and verification feedback~\cite{huang2024towards, sami2024aivril, blocklove2025automatically, wang2025veridispatcher}. Together, these developments mark a clear shift toward AI-assisted hardware design with the potential to accelerate development cycles and lower the barrier to digital system design.


\begin{figure}[h]
    \centering
    \includegraphics[width=0.5\textwidth]{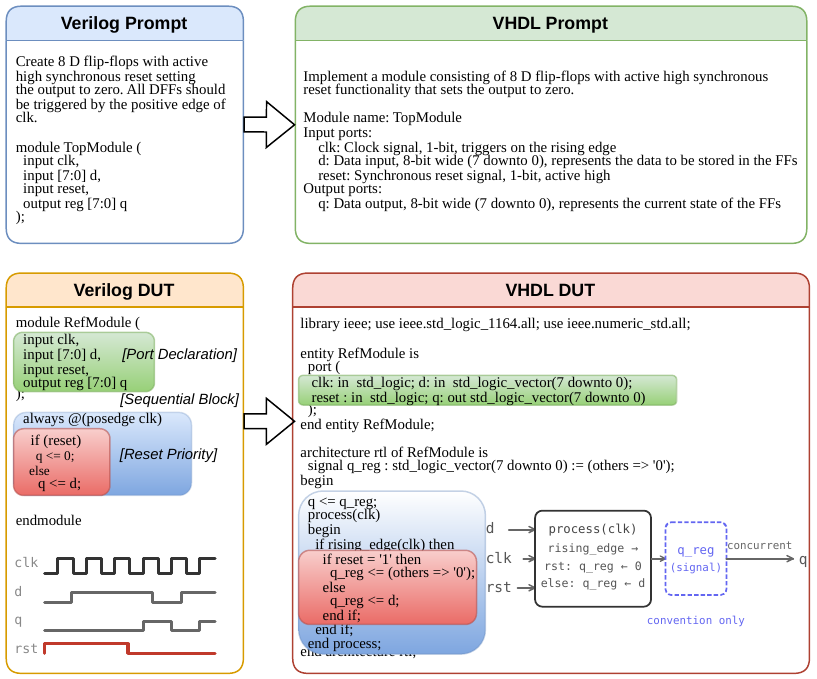}
    \caption{A case shows VHDLSuite convert Verilog to VHDL.}
    \vspace{-5mm}
\end{figure}

Despite this progress, most benchmarks and evaluations target Verilog~\cite{liu2024openllm,liu2023verilogeval}, leaving LLM capabilities on other HDLs less characterized. Evaluating models on structurally different languages is valuable because cross-lingual studies for LLMs have shown that training or fine-tuning on one programming language can transfer to others~\cite{baltaji2025cross}, underscoring the importance of benchmarking LLMs across related languages to fully capture their code generation capabilities. VHDL is a particularly informative complement to Verilog: where Verilog adopts a concise, C-like syntax and encapsulates designs within \texttt{module} blocks, VHDL follows a formal, Ada-inspired structure with stricter typing, explicit component declarations, and a separation of interface specifications (\texttt{entity}) from behavioral or structural descriptions (\texttt{architecture}) that requires explicit port mapping. These differences motivate dedicated VHDL evaluation, as Verilog-oriented benchmarks alone may not expose VHDL-specific typing, declaration, initialization, and structural error patterns.


Although early work has explored LLM-generated VHDL~\cite{vijayaraghavan2024vhdl, vijayaraghavan2024chain}, existing efforts face several limitations. Some rely on conversion tools such as the Icarus Verilog VHDL generator, which supports only a narrow range of constructs and often produces non-standard or non-synthesizable code, undermining evaluation validity. The field also lacks publicly available, rigorously validated VHDL benchmarks; existing datasets are typically small, proprietary, or missing complete testbenches for functional verification. Moreover, current evaluation methods often emphasize aggregate pass rates while providing limited insight into executable functional checking, syntactic validity, and VHDL-specific failure behavior. These gaps hinder reproducible research and limit a nuanced understanding of LLM performance on VHDL. To address these challenges, we introduce \texttt{VHDLSuite}, a benchmark-centered infrastructure that bootstraps scalable VHDL evaluation from mature Verilog benchmark corpora and enables reproducible analysis of LLM-generated VHDL. Our contributions are as follows:


\begin{itemize}[leftmargin=*]
\item \textbf{Data Pipeline.} An automated benchmark synthesis pipeline that converts Verilog designs and testbenches into executable VHDL benchmark instances with  VUnit/GHDL simulation verification.
\item \textbf{VHDLBench.} A benchmark of over 200 VHDL problems with validated testbenches spanning diverse difficulty levels.
\item \textbf{Comprehensive Evaluation.} A systematic assessment of state-of-the-art LLMs on VHDL generation, revealing key challenges and performance disparities that inform future multi-language hardware design automation.
\end{itemize}


\section{Background}

\subsection{Hardware Description Language (HDL)}
Hardware Description Languages (HDLs) are domain-specific languages for modeling concurrency and timing behavior of digital circuits ~\cite{chu2006rtl}. Verilog and VHDL are the two primary industry standards~\cite{berman1994standard}, but they embody fundamentally different design philosophies. Verilog, influenced by C, prioritizes conciseness~\cite{navabi1999verilog}, whereas VHDL, rooted in Ada, emphasizes verbosity and structural formalism~\cite{shahdad1986overview}. A defining characteristic of VHDL is its strict semantic rules and strong typing system~\cite{smith1996vhdl, ashenden2002vhdl}: unlike Verilog, it prohibits implicit type conversions and enforces a formal separation between the interface and logic architecture~\cite{roth2008digital}. While this rigidity enhances reliability in safety-critical domains such as aerospace~\cite{chu2006rtl}, it also means that traditional transpilers often fail to accurately map Verilog's loose syntax to VHDL's rigorous standards~\cite{ranathunga2023neural,li2000hml}, and the same structural differences pose substantial challenges for automated code generation.

\subsection{LLM for RTL Generation}

The application of LLMs in RTL design has evolved from basic code completion to sophisticated, multi-stage workflows~\cite{shao2024survey, wang2024llms, alsaqer2024potential}. Early milestones such as VeriGen~\cite{thakur2024verigen} and RTLCoder~\cite{liu2024rtlcoder} established the effectiveness of domain-specific fine-tuning, demonstrating that hardware-centric training data is essential for generating synthesizable RTL. Building on this foundation, frameworks like VeriCoder~\cite{wei2025vericoder} and ScaleRTL~\cite{deng2025scalertl} introduced iterative validation loops and reasoning-enhanced scaling to bridge the gap between initial synthesis and functional reliability~\cite{li2025eda}. These advances are complemented by standardized evaluation platforms such as RTLLM (OpenLLM-RTL)~\cite{liu2024openllm}, which provide functionally verified tasks that assess whether generated RTL meets design specifications rather than merely checking syntax~\cite{akyash2024evolutionary}. As recent reviews emphasize~\cite{abdollahi2024hardware}, such verified datasets and evaluation infrastructures are critical to the progress of LLM-based hardware design automation~\cite{wang2024llms}.

However, most of these efforts have focused on Verilog, with VHDL receiving less attention. Recent studies confirm that VHDL's verbosity and long-range structural dependencies lead to higher error rates in LLM-generated code compared to Verilog~\cite{vijayaraghavan2024vhdl, vijayaraghavan2024chain}. VHDL-Eval~\cite{vijayaraghavan2024vhdl} has initiated evaluation in this area but is limited by a small problem set and lacks validation for synthesis and resource usage. A large-scale, functionally verified benchmark is therefore needed to properly evaluate LLMs on VHDL code generation.


\section{Methods}

VHDLSuite is a benchmark construction and evaluation pipeline for deriving simulation-validated VHDL tasks from existing Verilog benchmarks and assessing LLMs on VHDL generation. The pipeline consists of four stages: source data standardization, LLM-based Verilog-to-VHDL translation, simulation-driven iterative verification, and unified evaluation. VHDLSuite separates benchmark construction from model evaluation. During construction, simulation feedback is used to validate and refine translated VHDL tasks. During evaluation, models generate candidates in a no-feedback setting, while the same VUnit/GHDL checker is used after generation to ensure comparable pass@k measurement across models.








\subsection{Data Pipeline}



\subsubsection{\textbf{Preprocessing}}

VHDLSuite draws its source data from existing Verilog benchmarks, including RTLLM and VerilogEval, which provide Verilog designs alongside natural language descriptions and executable testbenches, but differ in file organization, naming conventions, and description granularity, making direct cross-benchmark translation impractical. We therefore first standardize each problem instance into a uniform format by extracting three components: (1) the natural language description of the design's functionality, (2) the Verilog DUT code, and (3) the corresponding Verilog testbench. This standardization establishes a consistent input interface for subsequent translation and verification.


\subsubsection{\textbf{Code Translation}}
We use LLMs to translate the standardized Verilog designs and testbenches into VHDL design-testbench pairs, which are then validated as executable benchmark instances through simulation. To reduce HDL-specific syntactic violations, we incorporate a VHDL-aware system prompt that explicitly constrains design-unit structure, library usage, port declarations, and common syntax pitfalls. Given that VHDL differs substantially from Verilog in type systems, syntax structures, and simulation semantics, single-round generation often fails to produce compilable and functionally correct results. The translation process is therefore designed as iterative from the outset: rather than assuming first-attempt correctness, we treat each translation as a multi-round generation task. Once a verified VHDL implementation is obtained, we use an LLM to construct a refined generation prompt based solely on the original Verilog module description and the verified VHDL output, explicitly excluding all testbench code to prevent information leakage and ensure a clean evaluation interface.



\begin{figure}[!t]
    \centering
    \includegraphics[width=0.5\textwidth]{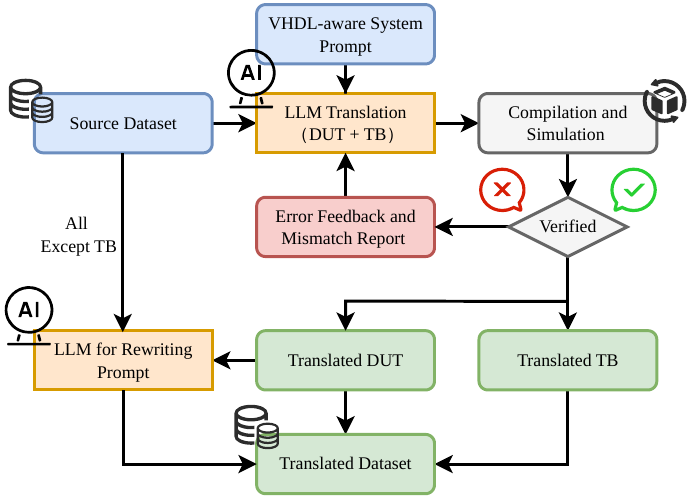}
    \caption{VHDLSuite data converting pipeline workflow.}
\end{figure}

\subsubsection{\textbf{Simulation and Repair driven Validation}}
\label{sec:verification}
Rather than directly accepting translated results, VHDLSuite subjects each generated VHDL design to an iterative, simulation-driven verification loop. Using VUnit, the pipeline automatically compiles and simulates translated VHDL DUTs against their VHDL testbenches, checking compilation correctness and executable behavioral consistency within the VHDL benchmark environment. When compilation failures or functional mismatches occur, VHDLSuite collects the corresponding error logs and feeds them back to the LLM along with the current code. In subsequent rounds, the model leverages both its conversation history and the simulation feedback to apply targeted corrections, progressively resolving type mismatches, port mapping errors, structural differences, and overlooked edge cases. Instances that fail to pass validation within the specified number of rounds are excluded from the final dataset, ensuring that all retained VHDLBench tasks compile successfully and pass their executable VHDL simulation tests. This construction criterion targets executable VHDL consistency: each retained instance is self-contained, simulation-checkable, and evaluated under the same VUnit/GHDL harness used for downstream model assessment.

\subsection{Benchmarking Framework}
\subsubsection{\textbf{Evaluation Loop}}
The evaluation framework reuses the compilation and VUnit simulation checker from Section~\ref{sec:verification}, but not the construction-time feedback loop. Each candidate is evaluated only after generation, so models do not receive simulation logs, reference implementations, Verilog sources, or testbench information during evaluation. All candidates are checked against the released VHDL testbench associated with each benchmark instance; the original Verilog artifacts are retained only for construction traceability and are never invoked by the evaluator. The key difference is that the evaluation stage imposes strict input constraints: unlike the construction-time loop, which accesses original Verilog DUTs and testbenches, it withholds all reference material to prevent information leakage. On this basis, VHDLSuite provides a unified, reproducible evaluation framework centered on simulation-driven verification, combining fine-grained error classification, multiple generation settings, and syntactic similarity measurement.



\paragraph{\textbf{Evaluation Metrics}}
Each generated VHDL output undergoes a phased verification process. First, outputs must compile; failures denote compilation errors from grammatical, type, or structural issues. Successfully compiled designs proceed to VUnit simulation; samples exhibiting unexpected functional behavior, including mismatches and timeouts, are classified as runtime errors. Only designs passing the full simulation are considered functionally correct. This two-tier failure classification enables detailed analysis of model capabilities at both syntactic and functional levels. To account for the stochasticity of LLM generation, we independently sample $n = 5$ outputs per problem and compute pass@1, pass@3, and pass@5, estimating the probability of obtaining at least one correct implementation within $k$ attempts. During evaluation, the model receives only the natural language problem description as input, lacking access to the testbench, Verilog reference, or simulation feedback.

\paragraph{\textbf{Automated Analysis}}
\label{para:structural_method}
To complement execution-based metrics, VHDLSuite incorporates a structural evaluation layer using the Tree-sitter parsing framework. Unlike traditional parsers, Tree-sitter is error-tolerant and produces Concrete Syntax Trees (CSTs) that retain detailed structural information even for partially incorrect code, enabling AST-like abstractions that quantify the syntactic similarity between generated designs and reference implementations. The similarity score is computed as a weighted mixture of sequential similarity ($S_{Seq}$) and Jaccard similarity ($S_{Jac}$), where a weight function $w(t)$ assigns higher priority to critical logic structures such as processes, control flows, and signal assignments over identifiers or literals, ensuring that the metric captures core design intent. Detailed mathematical definitions and specific weight configurations are provided in Section~\ref{sec:metrics_setup}.



\subsection{VHDLBench Dataset}


VHDLBench organizes all problem instances into a standardized file structure with a strict separation between construction-stage and evaluation-stage files. This design ensures reproducibility of functional verification while preventing leakage of original Verilog information during evaluation. Each problem can be independently compiled, simulated, and verified from its VHDL files alone, providing consistent correctness criteria for model benchmarking.

\subsubsection{\textbf{Source Benchmarks}}
VHDLBench draws from two Verilog benchmarks: \texttt{VerilogEval v2} \cite{liu2023verilogeval} and \texttt{RTLLM v2} \cite{lu2024rtllm}. VerilogEval focuses on problems ranging from simple to moderate difficulty, originating primarily from exercises on the HDLBits educational website. It tests understanding of specific grammar constructs, small modules, and state machines, making it well suited for evaluating foundational capabilities. RTLLM, in contrast, targets complete Intellectual Property core designs of industrial complexity, examining the ability to generate full functional modules involving substantial grammatical and logical challenges. Combining these two sources allows VHDLBench to cover a broad spectrum of translated VHDL generation scenarios, from basic constructs to industrial-scale designs, while leveraging mature Verilog corpora to bootstrap scalable VHDL evaluation in a resource-scarce benchmark setting.



\subsubsection{\textbf{Dataset Composition}}
VHDLBench comprises 206 problems: 156 derived from \texttt{VerilogEval v2} and 50 from \texttt{RTLLM v2}. Each problem is converted into a unified VHDL benchmark format through the construction pipeline described above. Every problem provides a self-contained set of files:


\begin{itemize}[leftmargin=*]
    \item \texttt{prompt.txt}: natural language problem description.
    \item \texttt{declaration.txt}: library and package dependencies.
    \item \texttt{dut.vhd}: VHDL Device Under Test.
    \item \texttt{tb.vhd}: VHDL testbench for simulation verification.
    \item \texttt{ref.vhd} (optional): Reference VHDL as golden output.
    \item \texttt{gen.vhd} (optional): Auxiliary stimulus/signal generator.
\end{itemize}

The presence of \texttt{ref.vhd} and \texttt{gen.vhd} varies by source benchmark. VerilogEval-derived problems separate stimulus generation from result checking and require an explicit reference design for golden comparison, whereas RTLLM-derived problems use self-contained testbenches that integrate both stimulus generation and correctness checking without a separate reference module. For construction-time traceability, we retain each problem's original Verilog files, but these are never exposed in the evaluation interface.



\section{Experiment Setup}
\subsubsection{\textbf{VHDL Simulation and Verification Framework}} 

We use VUnit, an open-source automated testing framework for VHDL that supports test execution, result collection, and assertion-based verification across mainstream simulators. VUnit serves as a Python-based verification interface that automatically compiles and simulates generated VHDL designs using GHDL as the backend.

\subsubsection{\textbf{Model Selection}}
We evaluate seven LLMs on VHDLBench: \texttt{Claude Sonnet 4.5}, \texttt{DeepSeek V3.2 Speciale}, \texttt{Gemini 3 Pro Preview}, \texttt{GLM 4.6V}, \texttt{GPT 5.1 Codex Max}, \texttt{Grok 4}, and \texttt{Qwen3 Max}. These models span diverse architectures, training data scales, and inference strategies, providing broad coverage of current mainstream LLM capabilities. To ensure comparability, all models are evaluated with identical inference hyperparameters: temperature $= 0.85$, top\_p $= 0.95$, top\_k $= 50$.

\subsubsection{\textbf{Evaluation Metrics.}}
\label{sec:metrics_setup}

\paragraph{Functional Metrics}
Each candidate output undergoes the same compilation-simulation-verification pipeline. Only results that pass all stages are considered functionally correct. We report statistics by aggregating the mean proportions of functionally correct results, compilation-only successes, and compilation failures, characterizing model capabilities in both VHDL syntax compliance and functional logic implementation. At the task level, we compute Pass@k ($k = 1, 3, 5$) to estimate the probability that at least one functionally correct VHDL implementation is obtained within $k$ independent generation attempts, capturing the stability of each model's generation behavior. Together, these metrics assess model performance from both outcome-level and task-level perspectives under consistent simulation-based verification.



\paragraph{Syntactic Similarity}
As introduced in the Section~\ref{para:structural_method}, we employ a custom weighted scoring system based on Tree-sitter-vhdl. We extract a serialized sequence of AST node types $T = \{t_1, t_2, \dots, t_n\}$ and define a weight function $w(t)$ that prioritizes logic-heavy constructs (e.g., $w(\text{process})=4.0$, $w(\text{assign})=3.0$) over common identifiers ($w(\text{id})=0.4$). The total structural score is:
\begin{equation}
Score = 0.8 \cdot S_{Seq} + 0.2 \cdot S_{Jac}
\end{equation}
where the two components are defined as follows. The \textbf{Weighted Sequence Ratio} ($S_{Seq}$) uses weighted matching based on the Gestalt Pattern Matching algorithm. Given matching blocks $M$ between reference and generated sequences:
\begin{equation}
\label{eq:score_formula}
S_{Seq} = \frac{2 \cdot \sum_{t \in M} w(t)}{\sum_{t \in T_{ref}} w(t) + \sum_{t \in T_{gen}} w(t)}
\end{equation}
The \textbf{Weighted Jaccard Similarity} ($S_{Jac}$) measures the overlap of node type distributions, where $C(k)$ denotes the count of node type $k$ in a sequence:
\begin{equation}
S_{Jac} = \frac{\sum_{k \in T_{ref} \cup T_{gen}} w(k) \cdot \min(C_{ref}(k), C_{gen}(k))}{\sum_{k \in T_{ref} \cup T_{gen}} w(k) \cdot \max(C_{ref}(k), C_{gen}(k))}
\end{equation}





\section{Benchmarking Results}
\label{sec:results}

\begin{figure}[htbp]
    \centering
    \includegraphics[width=0.5\textwidth]{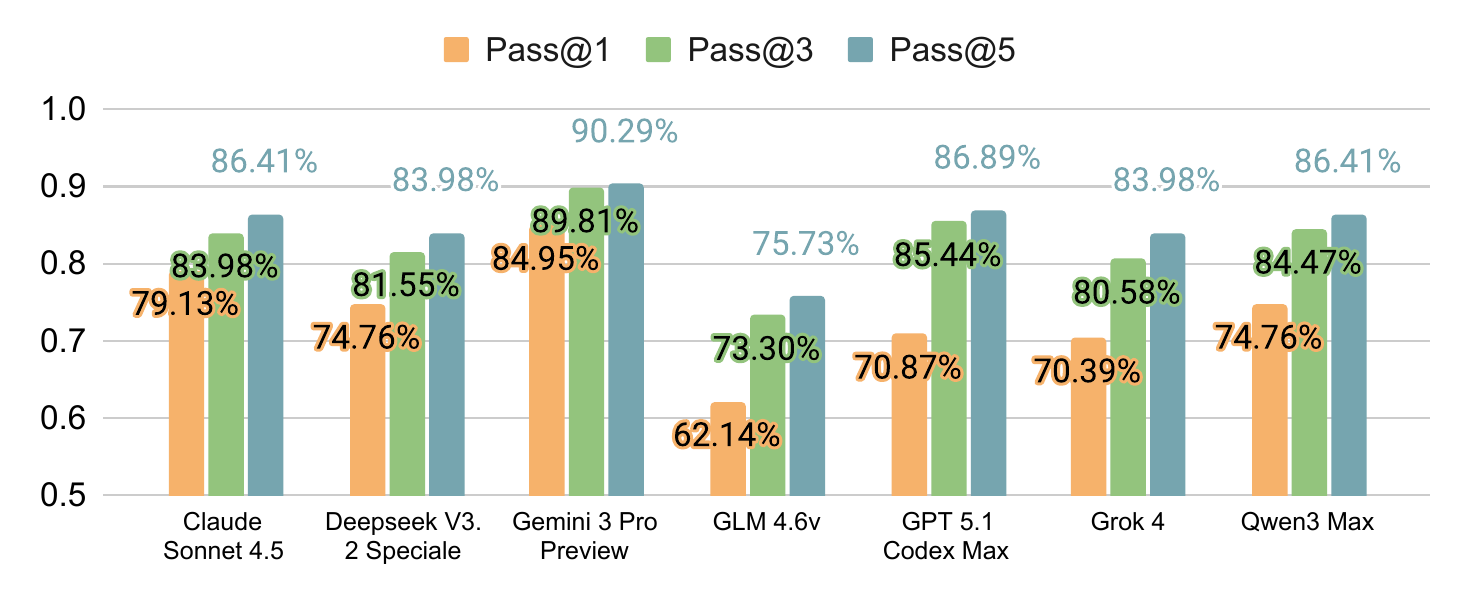}
    \caption{Pass@k Performance of LLMs on VHDLBench.}
    \label{fig:passk}
\end{figure}

Figure~\ref{fig:passk} presents pass@k performance across models. \texttt{Gemini 3 Pro Preview} achieves the highest scores at all $k$ values, while \texttt{GLM 4.6v} consistently ranks lowest. The remaining models form a tightly clustered group at pass@5 despite moderate differences at pass@1. Notably, \texttt{GPT 5.1 Codex Max} and \texttt{Grok 4} reach pass@5 results comparable to their peers but exhibit noticeably lower pass@1 scores, suggesting greater variability in single-sample reliability rather than a deficit in overall solution capability.

\begin{figure}[htbp]
    \centering
    \includegraphics[width=0.5\textwidth]{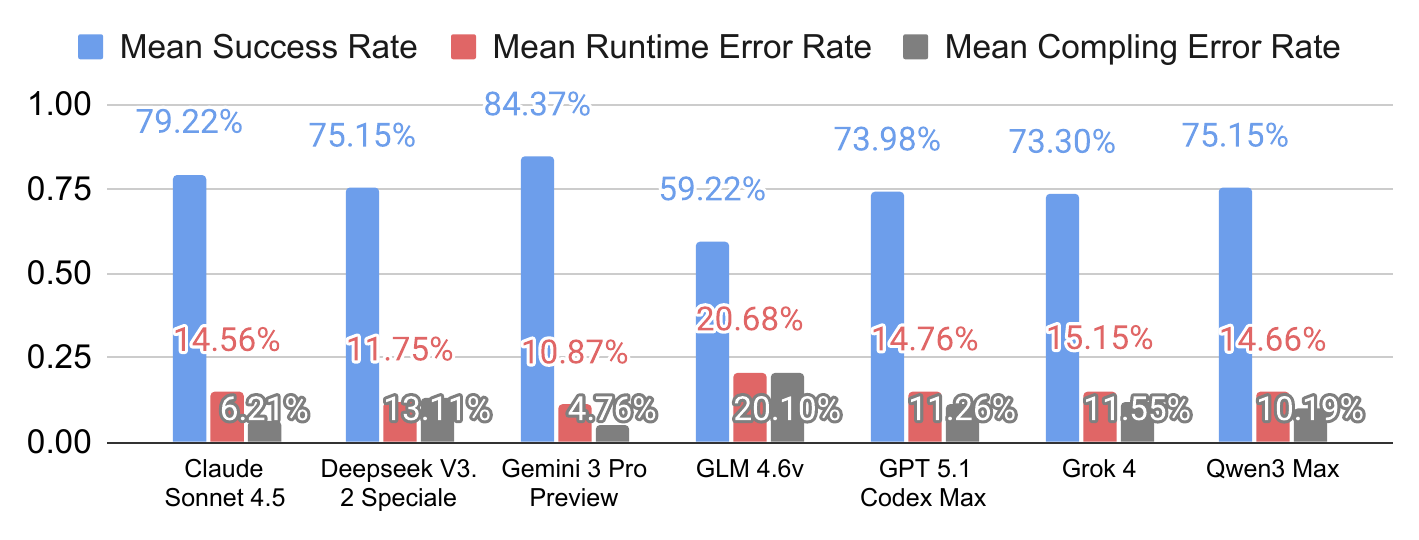}
    \caption{Outcome Distribution of Generated VHDL Designs.}
    \vspace{-6pt}
    \label{fig:outcome_dist}
\end{figure}

Figure~\ref{fig:outcome_dist} categorizes errors into runtime and compilation, revealing reliability differences not fully captured by pass@k. \texttt{Gemini 3 Pro Preview} again leads with the highest success rate and the lowest compilation error rate, indicating strong syntactic stability. \texttt{GLM 4.6v} shows the weakest profile, with reduced success and elevated error rates in both failure categories. Among the remaining models, \texttt{DeepSeek V3.2 Speciale} stands out with a comparatively higher compilation error rate despite a success rate on par with several peers, suggesting that its failures more frequently originate from structural or syntactic issues. \texttt{GPT 5.1 Codex Max} and \texttt{Grok 4} display more balanced error distributions, implying that their performance variations reflect generation stability rather than fundamental capability differences.

\section{Discussion}
\subsection{Generation Quality Evaluation}

Table~\ref{tab:ast_accuracy} shows AST-based syntactic similarity on functional successes, compilation failures, and runtime failures. A consistent trend holds across all models: successful designs exhibit the highest structural similarity, compilation failures the lowest, and runtime failures fall in between. As compilation failures break the AST logical sequence, sharply dropping $S_{Seq}$ (Eq.~\ref{eq:score_formula}), whereas runtime failures preserve much reference structure despite behavioral deviations.

\begin{table}[htbp]
  \centering
  \caption{AST Syntactic Similarity of Success and Failure Cases. }
  \label{tab:ast_accuracy}
  \footnotesize
  \resizebox{\columnwidth}{!}{
  \begin{tabular}{lcccccc}
    \toprule
    \multirow{2}{*}{Model} & \multicolumn{3}{c}{Translated RTLLM} & \multicolumn{3}{c}{Translated VerilogEval} \\
    \cmidrule(lr){2-4} \cmidrule(lr){5-7}
    & Success & Compiling & Runtime & Success & Compiling & Runtime \\
    \midrule
    Claude Sonnet 4.5      & \cellcolor{g2}82.95\% & \cellcolor{white}49.05\% & \cellcolor{g1}\textbf{70.04\%} & \cellcolor{g3}88.68\% & \cellcolor{g1}\textbf{76.70\%} & \cellcolor{g1}\textbf{83.58\%} \\
    DeepSeek V3.2 Speciale & \cellcolor{r3}81.41\% & \cellcolor{r1}\textbf{22.73\%} & \cellcolor{r2}59.99\% & \cellcolor{r3}87.42\% & \cellcolor{r1}\textbf{41.63\%} & \cellcolor{white}80.72\% \\
    Gemini 3 Pro Preview   & \cellcolor{r1}\textbf{78.99\%} & \cellcolor{r2}29.25\% & \cellcolor{r1}\textbf{57.92\%} & \cellcolor{r2}87.07\% & \cellcolor{g2}73.36\% & \cellcolor{r1}\textbf{77.90\%} \\
    GLM 4.6v               & \cellcolor{g1}\textbf{83.43\%} & \cellcolor{g1}\textbf{52.44\%} & \cellcolor{g2}66.80\% & \cellcolor{r1}\textbf{86.32\%} & \cellcolor{r2}61.83\% & \cellcolor{r2}78.13\% \\
    GPT 5.1 Codex Max      & \cellcolor{white}81.61\% & \cellcolor{g2}50.86\% & \cellcolor{white}64.32\% & \cellcolor{g1}\textbf{88.96\%} & \cellcolor{white}67.28\% & \cellcolor{r3}78.26\% \\
    Grok 4                 & \cellcolor{g3}82.25\% & \cellcolor{g3}50.86\% & \cellcolor{r3}60.75\% & \cellcolor{g2}88.94\% & \cellcolor{g3}69.28\% & \cellcolor{g3}81.14\% \\
    Qwen3 Max              & \cellcolor{r2}80.71\% & \cellcolor{r3}38.76\% & \cellcolor{g3}64.42\% & \cellcolor{white}87.95\% & \cellcolor{r3}63.16\% & \cellcolor{g2}82.40\% \\
    \bottomrule
  \end{tabular}
  }
\par \smallskip
{\footnotesize\textit{Note:} Cell shading indicates within-column rank. Green: top-3 (darker = better); Red: bottom-3 (darker = worse). \textbf{Bold} marks extremes per column.}
  
\end{table}
Average syntactic similarity for compilation-error cases is consistently higher across benchmarks on Translated VerilogEval than on Translated RTLLM, reflecting VerilogEval's more regular structural patterns relative to RTLLM's greater diversity and complexity. Notably, certain models such as \texttt{GLM 4.6v} retain relatively high similarity scores even under failure conditions, suggesting that their errors stem from localized semantic inconsistencies rather than large-scale structural distortion. Overall, these results confirm that functional correctness and structural similarity are correlated but not equivalent, underscoring the distinction between syntactic plausibility and execution validity in LLM-generated VHDL.


\subsection{Failure Type Analysis}

\begin{figure}[h]
    \centering
    \includegraphics[width=0.5\textwidth]{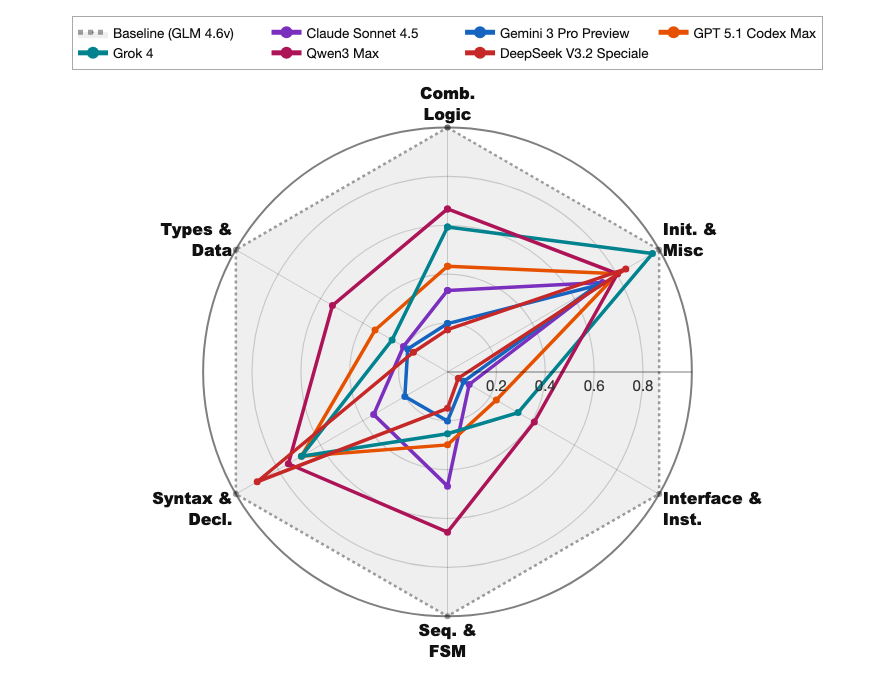}
    \caption{Topological Robustness Comparison: Performance across LLMs compared with \texttt{GLM 4.6v} Baseline on VHDL multiple Dimensions. 
    }
    \label{fig:error_radar}
    \vspace{-2pt}
\end{figure}

To analyze model failures beyond binary correctness, we introduce an LLM-driven failure taxonomy construction process. The LLM is prompted with the current error-type pool, the buggy design, and the golden DUT, and is constrained to assign at most three error labels per failure instance. When a failure does not match any existing label, new labels are dynamically appended to the pool, enabling incremental refinement. The resulting taxonomy is organized into six categories capturing distinct structural and semantic failure modes: \textit{Syntax \& Declaration}, \textit{Types \& Data Objects}, \textit{Interfaces \& Instantiation}, \textit{Combinational Logic}, \textit{Sequential \& FSM}, and \textit{Initialization \& Miscellaneous}. Figure~\ref{fig:error_radar} visualizes category-wise normalized error ratios computed as:
\begin{equation}
    R_{m,c} = \frac{E_{m, c}}{\max_{m'}E_{m',c}}
\end{equation}
where $E_{m,c}$ denotes the number of errors from model $m$ in category $c$. Since \texttt{GLM 4.6v} consistently exhibits the largest normalized ratios across categories, it serves as a natural empirical baseline for comparison. A clear pattern emerges: models with stronger overall functional performance exhibit uniformly lower error ratios across nearly all categories. Rather than revealing category-level trade-offs, Figure ~\ref{fig:error_radar} shows that performance differences stem from a global reliability factor, with higher-performing models demonstrating broadly reduced error magnitudes across all failure modes.

\begin{figure}[h]
    \centering
    \caption{Cross-category Error Distribution: Global Top 5 Failure Modes with Hierarchical Taxonomy Prefixes. 
    }
    \includegraphics[width=0.5\textwidth]{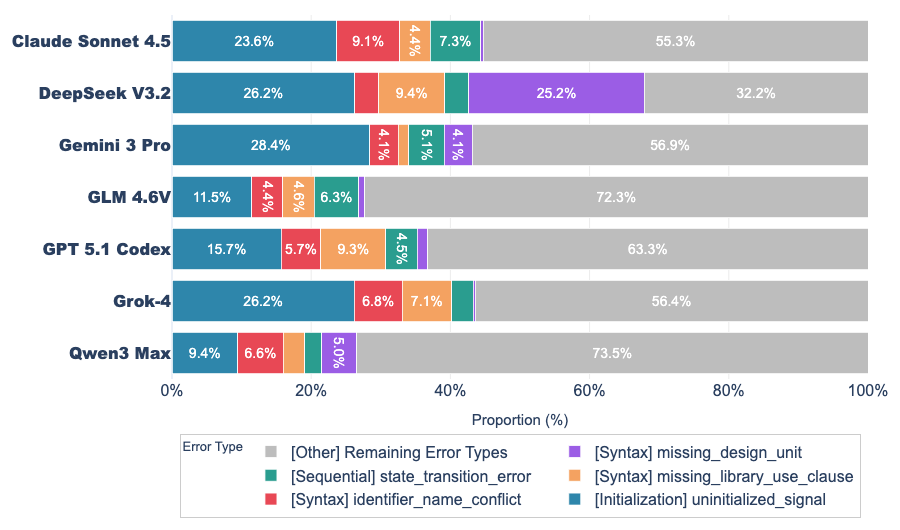}
    \label{fig:global_error}
\end{figure}

Figure~\ref{fig:global_error} presents dominant error category distributions across all evaluated models. The most widespread failure type is \texttt{uninitialized}\\\texttt{\_signal}, typically caused by incomplete handling of VHDL's signal initialization semantics. A similar pattern holds for \texttt{identifier\_name}\\\texttt{\_conflict} errors, reflecting VHDL-specific naming constraints. These prevalent errors suggest that, despite strong functional performance, many models rely on generalized HDL knowledge without fully internalizing VHDL-specific conventions.


A distinct outlier is \texttt{DeepSeek V3.2 Speciale}, which exhibits a disproportionately large number of \texttt{missing\_design\_unit} errors, accounting for a substantial share of its compilation failures. Manual inspection reveals that many of these cases correspond to incomplete outputs where the design unit is entirely absent. This behavior is plausibly linked to extended reasoning processes that fail to complete within the generation window, resulting in truncated or empty responses. The pattern thus appears to reflect limitations in sustaining coherent generation over structurally demanding tasks rather than isolated syntactic misunderstandings.




\subsection{Difficulty and Conversion Success Rate}
\label{sec:manual_inspection}

To characterize the difficulty distribution of VHDLBench, we classify each problem by its average pass@1 rate across all seven evaluated models, defining three soluble tiers (\textit{easy}, \textit{medium}, and \textit{hard}) along with an \textit{insoluble} category for problems that no model succeeded in any attempts. Table~\ref{tab:difficulty} summarizes the results distribution.


\begin{table}[h]
\centering
\caption{Difficulty Distribution of VHDLBench Based on Average Pass@1 Across All Models}
\label{tab:difficulty}
\small
\setlength{\tabcolsep}{4pt}
\begin{tabular}{lcccr}
\toprule
\textbf{Difficulty} & \textbf{Pass@1 Range} & \textbf{VerilogEval} & \textbf{RTLLM} & \textbf{Total (\%)} \\
\midrule
Easy       & $>80\%$              & 99  & 24 & 123 (59.7) \\
Medium     & $40\text{--}80\%$    & 28  & 14 & 42\ \ (20.4) \\
Hard       & $0\text{--}40\%$     & 23  & 10 & 33\ \ (16.0) \\
Insoluble  & $0\%$ (all models)   & 6   & 2  & 8\ \ \ \ (3.9) \\
\midrule
\textbf{Total} &                  & \textbf{156} & \textbf{50} & \textbf{206 (100)} \\
\bottomrule
\end{tabular}
\end{table}


Manual inspection of the eight insoluble problems reveals that their failures stem from systematic degradation during Verilog-to-VHDL prompt translation rather than inherent task difficulty. These cases are not concentrated in RTLLM: six come from VerilogEval and two from RTLLM, suggesting unsolvability is not simply tied to design complexity. Two recurring patterns are identified. First, structured information originally conveyed through state diagrams, Karnaugh maps, and timing waveforms was replaced with prose that failed to preserve the underlying logical constraints, leaving prompts fundamentally underspecified. Second, several translations introduced outright functional errors in port interfaces or control signal semantics, rendering a testbench-passing solution impossible regardless of model capability. Beyond these failure-inducing errors, inspection reveals a broader tendency for translated prompts to over-prescribe implementation details such as state counts, internal signal names, and pipeline structures. While this does not directly cause failures, it may lower the effective difficulty of problems originally designed to test a model's ability to derive correct designs from high-level specifications, potentially compressing the difficulty distribution and reducing the benchmark's discriminability.


\subsection{Case Studies}
\label{sec:case_studies}
We show a representative failure mode arising from VHDL's compilation semantics using a 16-bit ripple-carry full adder (\texttt{adder\_16bit}) as a case study. This example highlights a common misalignment between LLM-generated code and VHDL's strict requirements, while demonstrating the pipeline's automated correction through structured error feedback. This case illustrates the role of construction-time simulation feedback while showing that many VHDL failures stem from language-semantics issues rather than high-level design-intent errors. The module uses a four-level compositional hierarchy: a 1-bit full adder (\texttt{add1}) is composed into 2-bit, 4-bit, 8-bit, and finally the 16-bit top-level entity, as shown in Figure~\ref{fig:case_study}.

\begin{figure}[htbp]
\vspace{-10pt} 
\centering
\begin{minipage}{0.97\linewidth}

\pipelineblock{Prompt (task description)}{blue}{
  Implement a module of a 16-bit full adder in combinational logic using a hierarchical ripple-carry architecture...
}

\vspace{-6pt} 

\pipelineblock{LLM Generated (Round 1)}{red}{
  entity adder\_16bit is ...   \textcolor{red}{$\leftarrow$ declared first}\\
  \hspace*{1em} add8\_inst2 : entity work.add8 ...\\
  \hspace*{1em} add8\_inst1 : entity work.add8 ...\\
  entity add8 is ...           \textcolor{red}{$\leftarrow$ add8 defined AFTER use}\\
  ...\\
  entity add1 is ...           \textcolor{red}{$\leftarrow$ base case at bottom}
  
  \vspace{2pt}
  \hrule
  \vspace{2pt}
  
  {\fontsize{7.5pt}{8.5pt}\selectfont \color{red!80!black}
  error: unit "add8" not found in library "lib"\\
  \hspace*{1em} add8\_inst2 : entity work.add8\\
  \hspace*{1em} add8\_inst1 : entity work.add8\\
  error: entity "add8" was not analysed\\
  error: entity "add4" was not analysed\\
  ... (6 errors total) --- \textbf{Compile failed}}
}

\vspace{-6pt}

\pipelineblock{LLM Corrected (Round 2)}{green!60!black}{
  entity add1 is ...           \textcolor{green!70!black}{$\leftarrow$ base case first}\\
  entity add2 is ...           \textcolor{green!70!black}{$\leftarrow$ bottom-up order}\\
  entity add4 is ...\\
  entity add8 is ...\\
  entity adder\_16bit is ...   \textcolor{green!70!black}{$\leftarrow$ top-level last}\\
  
  \vspace{2pt}
  \hrule
  \vspace{2pt}
  
  {\fontsize{7pt}{8pt}\selectfont \color{black!70}
  Hint: Output 'y' has no mismatches.\\
    Hint: Output 'Co' has no mismatches.\\
    Hint: Total mismatched samples is 0 out of 100 samples\\
    Simulation finished at 1030000 ps\\
    \textbf{Mismatches: 0 in 100 samples}}
}

\end{minipage}
\caption{Example on a hierarchical adder case. Each block displays the generated VHDL code (top) and the corresponding simulator feedback (bottom). }
\vspace{-2mm}
\label{fig:case_study}
\end{figure}
The LLM produced logically correct VHDL with proper port mappings and carry logic, but declared entities in a top-down order. While valid in Verilog, this violates VHDL's sequential analysis requirement: an entity must be compiled before any higher-level module can instantiate it, making bottom-up ordering mandatory. The top-down ordering triggered six compilation errors in the first round, which VHDLSuite captured and returned as structured feedback. Specifically, \texttt{work.add8} was instantiated before \texttt{add8} was declared, causing unresolved design-unit errors despite correct carry logic. This error reflects an imbalance in current LLMs' HDL knowledge: their stronger familiarity with Verilog can preserve correct RTL composition, but may make them less sensitive to VHDL's stricter declaration-order requirements during analysis. In the second round, the model reordered the declarations into a bottom-up sequence (\texttt{add1} $\to$ \texttt{adder\_16bit}) without altering the underlying logic, resolving all compilation issues. Subsequent functional simulation confirmed zero mismatches across 100 test vectors for both the \texttt{y} and \texttt{Co} outputs.

\section{Conclusion and Future Work}
We introduce VHDLSuite, a benchmark-centered infrastructure for constructing executable VHDL tasks and evaluating VHDL code generated by Large Language Models (LLMs) through simulation-based functional checking and diagnostic error analysis. To address the scarcity of VHDL evaluation resources, we present an automated data pipeline that translates Verilog designs and testbenches into executable, simulation-validated VHDL benchmark instances. Utilizing this pipeline, we developed VHDLBench, a benchmark comprising over 200 validated VHDL problems across varying complexity levels. Through extensive evaluation of state-of-the-art LLMs, we highlight critical VHDL-specific generation challenges and provide essential insights to advance multi-language hardware design automation. Building upon this foundation, future work will focus on leveraging LLMs to facilitate translations between a broader range of hardware description languages. Additionally, we aim to optimize token efficiency during the translation process and further enhance verification efficiency. Ultimately, these improvements will enable the automated synthesis and validation of large-scale, multi-language hardware datasets. Future work will extend the infrastructure beyond benchmark conversion, translating training data to strengthen models' VHDL capability while adding benchmarks for VHDL-specific syntax features relative to Verilog, such as packages, generics, records, and type conversions.

\bibliographystyle{ACM-Reference-Format}
\bibliography{refs}

\appendix
\appendix
\section{Verilog Baseline Evaluation and Comparison}

To contextualize VHDLBench results, we evaluate \texttt{Claude Sonnet 4.5}, \texttt{GPT 5.1 Codex Max}, and \texttt{GLM 4.6v} on the original Verilog versions of both source benchmarks. Results are summarized in Table~\ref{tab:verilog-baseline}.

\begin{table}[h]
\caption{Verilog Baseline Performance (n=5)}
\label{tab:verilog-baseline}
\begin{tabular}{lccc}
\toprule
Model & Success & Runtime & Compile \\
\midrule
\multicolumn{4}{l}{\textit{VerilogEval}} \\
Claude Sonnet 4.5 & 80.0\% & 16.9\% &  3.1\% \\
GPT 5.1 Codex Max & 87.3\% &  8.3\% &  4.4\% \\
GLM 4.6v          & 68.6\% & 13.3\% & 18.1\% \\
\midrule
\multicolumn{4}{l}{\textit{RTLLM}} \\
Claude Sonnet 4.5 & 53.2\% & 18.4\% & 28.4\% \\
GPT 5.1 Codex Max & 55.2\% & 17.2\% & 27.6\% \\
GLM 4.6v          & 40.4\% & 11.2\% & 48.4\% \\
\bottomrule
\end{tabular}
\end{table}

Two observations follow from this comparison. First, the relative difficulty ordering between VerilogEval and RTLLM is preserved after translation: all three models show consistently lower success rates and higher compilation error rates on RTLLM than on VerilogEval in both Verilog and VHDL, confirming that VHDLBench relatively faithfully reflects the intrinsic difficulty distribution of its source benchmarks rather than introducing artifacts through translation. 

Second, comparing these Verilog baselines against the VHDL results in Section~\ref{sec:results} reveals varying degrees of cross-lingual generalization across models. \texttt{GPT 5.1 Codex Max} performs notably stronger on Verilog than VHDL, while \texttt{Claude Sonnet 4.5} shows the opposite pattern, and \texttt{GLM 4.6v} degrades substantially in Verilog relative to VHDL. This inconsistency suggests that HDL generalization may be model-specific, possibly influenced by differences in HDL coverage across pretraining corpora, though further investigation is needed to substantiate this hypothesis.

\definecolor{cskeyword}{RGB}{0,0,180}
\definecolor{cscomment}{RGB}{0,128,0}
\definecolor{csbg}{RGB}{248,248,248}
\definecolor{csdel}{RGB}{180,0,0}
\definecolor{csadd}{RGB}{0,120,0}
 
\lstdefinelanguage{vhdlx}{
  morekeywords={library,use,entity,architecture,port,in,out,signal,
    begin,end,process,if,elsif,else,then,is,of,constant,type,array,
    rising_edge,downto,others,std_logic,std_logic_vector,record,variable,
    integer,wait,for,until},
  sensitive=false,
  morecomment=[l]{--},
  morestring=[b]",
}
 
\lstdefinestyle{vhdlstyle}{
  language=vhdlx,
  basicstyle=\ttfamily\scriptsize,
  keywordstyle=\color{cskeyword}\bfseries,
  commentstyle=\color{cscomment}\itshape,
  stringstyle=\color{black},
  backgroundcolor=\color{csbg},
  showstringspaces=false,
  columns=fullflexible,
  keepspaces=true,
  breaklines=true,
  frame=single,
  framerule=0.3pt,
  xleftmargin=4pt,
  xrightmargin=2pt,
  aboveskip=4pt,
  belowskip=2pt,
}

\section{Case Study: Testbench-Induced Misjudgment on a Correct Design}
\label{app:jc_counter}

The case study in Section~\ref{sec:case_studies} shows a failure rooted
in the \emph{generated design}: logically correct carry logic that GHDL
rejects because of VHDL's declaration-order semantics, then repaired
through construction-time feedback. Here we present the opposite case,
where the design under test (DUT) is correct from the first round and
the repair iterations are spent entirely on the
\emph{generated testbench}. The example is \texttt{JC\_counter} from
RTLLM~v2 (\texttt{Control/Counter}), a 64-bit Johnson (twisted-ring)
counter. In the construction-time repair loop
(Section~\ref{sec:verification}), this instance took four rounds before
it passed validation, which at first glance looks like a design the
model had trouble translating. The round-by-round trace tells a
different story.

\subsection*{The DUT Is Correct and Unchanged}
The generated DUT is logically identical across all four rounds; the
last round differs from the earlier ones only in removed comments. The
core process is a textbook Johnson recurrence: on each rising clock edge
it reads the least-significant bit \texttt{Q\_reg(0)}, shifts right, and
writes the complement of that bit into the most-significant position.

\begin{lstlisting}[style=vhdlstyle,caption={Generated DUT process, unchanged across all four rounds (comments elided).},label={lst:jc-dut}]
johnson_proc : process(clk, rst_n)
begin
  if rst_n = '0' then
    Q_reg <= (others => '0');
  elsif rising_edge(clk) then
    if Q_reg(0) = '0' then
      Q_reg <= '1' & Q_reg(63 downto 1);
    else
      Q_reg <= '0' & Q_reg(63 downto 1);
    end if;
  end if;
end process;
Q <= Q_reg;
\end{lstlisting}

Tracing the recurrence by hand from the all-zero reset state, the value
after 20 clock cycles is \texttt{0xFFFFF00000000000}, with the high 20
bits set, which is the expected point in the Johnson sequence. The DUT
is therefore already correct; the iterations did not change it, and the
final round passes all 132 samples. What the repair loop actually fixed
was the testbench that judged the design.

\subsection*{Round-by-Round Evolution}
Table~\ref{tab:jc-rounds} traces the four rounds. The mismatch count
falls from four to zero without any change to the DUT, and the two
substantive edits both land in the testbench.

\begin{table}[th]
\centering
\caption{Repair-loop evolution for \texttt{JC\_counter}. Every mismatch
originates in the generated testbench. The ``Mism.'' column reports
mismatches over total samples, followed by the time of the first
mismatch (\texttt{@}).}
\label{tab:jc-rounds}
\footnotesize
\setlength{\tabcolsep}{3pt}
\begin{tabularx}{\linewidth}{@{}l l Y Y@{}}
\toprule
Round & Mism. (@ first) & TB change & Effect \\
\midrule
v0 & 4/136 @ 705\,ns & ---             & redundant \texttt{wait} + wrong \texttt{exp[0]} \\
v1 & 1/132 @ 0       & remove waits    & wrong \texttt{exp[0]} remains \\
v2 & 1/132 @ 0       & reword comments & no change \\
v3 & \textbf{0/132} @ --- & fix \texttt{exp[0]} & passes \\
\bottomrule
\end{tabularx}
\end{table}

\paragraph{Sampling alignment (v0$\rightarrow$v1).}
The round-0 stimulus reaches each checkpoint with
\texttt{wait for PERIOD*N} and then issues an additional
\texttt{wait until rising\_edge(clk)} before raising the check enable.
Removing these redundant waits in round~1 changes which clock edge the
checker samples on, and the mismatch count drops from four to one. The
edits are confined to the stimulus timing; the DUT is untouched, so the
remaining mismatch cannot be a design error.

\paragraph{Miscomputed golden value (v2$\rightarrow$v3).}
The last mismatch comes from a hard-coded expected constant that the
model got wrong when it wrote the testbench.

\begin{lstlisting}[style=vhdlstyle,caption={The only substantive change between v2 and v3: a corrected golden constant.},label={lst:jc-diff}]
-- v0--v2 (wrong)
0 => x"FFFFFFE000000000",
-- v3 (correct): high 20 bits (63 downto 44) set
0 => x"FFFFF00000000000",
\end{lstlisting}

The correct value \texttt{0xFFFFF00000000000} sets the high 20 bits, as
the hand trace shows; \texttt{0xFFFFFFE000000000} sets the wrong bits.
Round~2 only reworded timing comments and reproduced the round-1 result.
Round~3 corrected the constant, left the DUT alone, and the design
passed all 132 samples.

\subsection*{Discussion}
This case complements the design-level failure in
Section~\ref{sec:case_studies} by exposing a construction-stage hazard
on the verification side. When one LLM writes both the DUT and a
self-checking testbench, the testbench brings its own sources of
misjudgment that are independent of the design: stimulus timing that
samples on the wrong edge, and hard-coded golden values that the model
derives incorrectly. The risk is sharper for RTLLM-style problems, whose
self-contained testbenches fold stimulus generation and result checking
into a single module with no separate reference, and there is thus no golden
artifact to cross-check the embedded expected values. With a smaller
repair budget, this instance would have been dropped as non-convertible
during construction even though its design was correct. This extends the
observation in Section~\ref{sec:manual_inspection} that apparent
unsolvability often traces back to the construction-stage flaws and model hallucinations rather than intrinsic task difficulty. Whereas that previous analysis centers on the stimulus and specification side, the present case exposes the same effect on the verification side, which argues for keeping DUT correctness and testbench correctness separate during construction. One partial remedy for
RTLLM-derived testbenches is to drive checking from an algorithmic
reference process instead of manually enumerated expected values, though
this helps little for designs that lack a concise independent reference,
such as complex finite-state machines.

\end{document}